\documentclass[10pt,letterpaper]{article}
\usepackage{opex3}
\usepackage{graphicx}
\usepackage{amsmath}
\usepackage{amssymb}
\usepackage{hyperref}
\usepackage{epstopdf}
\usepackage{color}
\usepackage{cite}
\begin{document}
\title{Generation of intense circularly polarized attosecond light bursts from relativistic laser plasmas}
\author{Guangjin Ma$^{1,*}$, Wei Yu$^{1}$, M. Y. Yu$^{2,3}$, Baifei Shen$^1$ and Laszlo Veisz$^{4}$}

\address{$^1$State Key Labratory of High Field Laser Physics, Institute of Optics and Fine Mechanics, Chinese Academy of Sciences, Shanghai 201800, China\\
$^2$Institute of Fusion Theory and Simulation, Zhejiang University, Hangzhou 310027, China\\
$^3$Institute for Theoretical Physics I, Ruhr University, Bochum, D-44780 Germany\\
$^4$Max-Planck-Institut f\"ur Quantenoptik, D-85748 Garching, Germany}
\email{$^*$guangjin@siom.ac.cn}

\begin{abstract}
We have investigated the polarization of attosecond light bursts generated by nanobunches of electrons from relativistic few-cycle laser pulse interaction with the surface of overdense plasmas. Particle-in-cell simulation shows that the polarization state of the generated attosecond burst depends on the incident-pulse polarization, duration, carrier envelope phase, as well as the plasma scale length. Through laser and plasma parameter control, without compromise of generation efficiency, a linearly polarized laser pulse with azimuth $\theta^i=10^\circ$ can generate an elliptically polarized attosecond burst with azimuth $|\theta^r_{\rm atto}|\approx61^\circ$ and ellipticity $\sigma^r_{\rm atto}\approx0.27$; while an elliptically polarized laser pulse with $\sigma^i\approx0.36$ can generate an almost circularly polarized attosecond burst with $\sigma^r_{\rm atto}\approx0.95$. The results propose a new way to a table-top circularly polarized XUV source as a probe with attosecond scale time resolution for many advanced applications.
\end{abstract}





\section{Introduction}
Compared to the high-order harmonic generation (HHG) from noble gases, HHG from plasma surfaces\cite{rmp81.445Teubner09} does not subject to the limitation of maximum applied laser intensity and can thus use the state-of-the-art terawatt and petawatt laser technology. Since the availability of powerful lasers with intensity-wavelength squared $>10^{16} {\rm W cm^{-2} \mu m^2}$, much work\cite{prl46.29Carman81,pra24.2649Carman81,apl31.172Burnett77, prl76.1832Norreys96, prl98.103902Tarasevitch07,prl110.175001Kahaly13, prl108.235003Heissler12, apb101.511Heissler10} has been done with different laser technologies. Currently, measurements with controlled laser and plasma parameters\cite{prl110.175001Kahaly13} on HHG has found remarkable spatial features\cite{nphys5.146Dromey09}, high brightness \cite{prl99.085001Dromey07}, as well as attosecond (as) duration\cite{nphys5.124Nomura09}, making it potentially useful to pump-probe experiments\cite{rmp81.163Krausz09}.

On the other hand, circularly polarized light in the extreme ultra violet (XUV) and soft x-ray regions has proven to be very useful for applications including the direct measurement of quantum phases in graphene and topological insulators\cite{nphys8.616Xu12,nanolett12.3900Gierz12}, the XUV magnetic circular dichroism spectroscopy\cite{pscr1993.302Schuetz93,pnas112.14206Fan15}, as well as the reconstrcution of band structure and modal phases in solids\cite{prl107.166803Liu11}. Currently, such radiation is mainly available at the large scale synchrotron sources and the time resolution is far below the sub-laser cycle time scale, limiting its wide availability and time resolving power. Although XUV optics in general can be used to convert XUV polarization\cite{oe23.033564Schmidt15}; its bandwidth is typically narrow and transmission efficiency is not high. There is strong motivation to directly realize a table-top circularly polarized attosecond XUV source by using laser. Some recent breakthroughs\cite{nphoton9.99Kfir15} has been made in HHG from noble gases. An attosecond XUV source by HHG from plasma surfaces has the potential to further improve attosecond burst energy. It is timely for us to investigate the polarization of harmonics generated from plasma surfaces and to expore how to efficiently generate a circularly polarized attosecond light burst in the XUV region.

Polarization of the harmonics is one of the basic properties regularily investigated in HHG from plasma surfaces. Many authors have considered the dependence of the harmonic intensity and polarization state on the incident laser pulse\cite{njp10.025025Rykovanov08, pre74.065401Baeva06,oc198.419Gal01, epl48.390Veres99,pop3.3425Lichters96, prl76.50Gibbon96, prl76.1832Norreys96, prl76.2278Gizzi96, prl112.123902Yeung14,apb101.511Heissler10, apb82.13Racz06,jqe28.2388Linde92}. Three mechanisms of harmonic generation, namely relativistic oscillating mirror (ROM)\cite{pop3.3425Lichters96,pre74.046404Baeva06}, coherent wake emission (CWE)\cite{prl96.125004Quere06}, and coherent synchrotron emission (CSE)\cite{pop17.033110anderBrugge10,prl109.245005Mikhailova12}, have been identified. With ROM, the high-order harmonics are purely doppler upshifted reflections from the oscillating plasma surface and their polarization more or less follows that of the incident laser. With CWE, because of linear mode conversion, only the $p-$polarized component of the laser field is effective, so that the resulting harmonics are also $p-$polarized and their intensity is proportional to the cosine of the polarization angle\cite{apb101.511Heissler10}. There exists still not much investigation on the polarization of the harmonics from the CSE mechanism. Yeung et al.\cite{prl112.123902Yeung14} experimentally investigated the polarization dependence of relativistic laser-driven high-harmonic emission from thin foils in the CSE context for the normal incidence in transmission geometry. The resulting polarization of the harmonics was found to be similar to that of the ROM harmonics.

In this paper we investigate the polarization of CSE harmonics for the oblique incidence in reflection geometry. The dependence of the polarization state of the attosecond light bursts on that of the driver laser pulses is investigated for the first time to the best of our knowledge. Based on this study, we propose a new way to generate attosecond burst with circular polarization or at least with elliptical polariation and a high degree of ellipticity. The structure of the paper is organized as follows: In Sec.~\ref{Sec_repr}, we summarize the general representation of light polarization state. In Sec.~\ref{Sec_exam}, we discuss properties of attosecond light bursts in a typical interaction scenario. In Sec.~\ref{Sec_modl}, we analyze the energy coupling process revealing the underlying physics as proposed in our previous work\cite{pop22.033105Ma15}. Sec.~\ref{Sec_para} is to the study of the parametric dependence. At last, our work is concluded and its applicability is discussed in Sec.~\ref{Sec_conc}.

\section{Representation of light polarization}\label{Sec_repr}
In a right-handed coordinate system, a generally elliptically polarized light pulse propagating along $x$-axis is described by\cite{EP77Azzam} $\mathbf{E}_{\perp}(t)={\mathrm Re} \{ \tilde{E}(t) \exp[j\phi(t)]\mathbf{\hat{e}}_\rho \}= {\rm {Re}} \{ \tilde{E}_y(t) \exp[j\phi_y(t)]\mathbf{\hat{y}} \} + {\rm {Re}} \{ \tilde{E}_z(t) \exp[j\phi_z(t)] \mathbf{\hat{z}} \}$. Here $\tilde{E}$ and $\phi$ are the pulse ampltitude and the absolute phase. $\mathbf{\hat{e}}_{\rho} = (\cos\theta \cos\epsilon - j \sin\theta \sin\epsilon)\mathbf{\hat{y}} + (\sin\theta \cos\epsilon + j \cos\theta \sin\epsilon)\mathbf{\hat{z}}$ is the polarization vector with $\theta$ the azimuth angle limited to the range $-\pi/2 \leq \theta < \pi/2$, and $\epsilon$ the ellipticity angle to the range $-\pi/4 \leq \epsilon \leq \pi/4$. For light pulse propagating in positive (negative) $x$-direction, $\epsilon>0$ corresponds to right-hand (left-hand), while $\epsilon<0$ corresponds to left-hand (right-hand) elliptical polarization, when looking against its propagation direction.

A Stokes vetor $\mathbf{S}=[S_0,S_1,S_2,S_3]$ is often used to describe light polarization state,
\begin{subequations}\label{E_stokes}
\begin{align}
S_0 &= <\tilde{E}^2_y(t)> +  <\tilde{E}^2_z(t)> \\
S_1 &= <\tilde{E}^2_y(t)> -  <\tilde{E}^2_z(t)> \\
S_2 &= <\tilde{E}_y(t)\tilde{E}_z(t)\cos[\phi_z(t)-\phi_y(t)]> \\
S_3 &= <\tilde{E}_y(t)\tilde{E}_z(t)\sin[\phi_z(t)-\phi_y(t)]>
\end{align}
\end{subequations}
where $<v>$ signifies the time average of $v$: $<v>=(1/T)\int_0^T v dt$, with $T$ an interval of time long enough to make the time-average integral independent of $T$ itself. The azimuth $\theta$ and ellipticity angle $\epsilon$ are calculable from these Stokes parameters,
\begin{subequations}\label{E_polang}
\begin{align}
\theta &= (1/2) \arctan(S_2/S_1) \\
\epsilon &= (1/2) \arcsin[S_3/(S^2_1+S^2_1+S^2_3)^{1/2}]
\end{align}
The ellipticity of the polarization ellipse $\sigma$ is related to $\epsilon$ through,
\begin{align}
\sigma = \tan\epsilon
\end{align}
\end{subequations}
It is limited to the range $-1\leq \sigma \leq 1$.

An alternative description of light polarization is through a complex parameter $\chi$, defined as: $\chi = (\tilde{E}_z/\tilde{E}_y) \exp{j[\phi_z - \phi_y]}$, it is connected to $\theta$ and $\epsilon$ through,
\begin{align}\label{E_polchi}
\chi = \frac{\tan\theta + j \tan\epsilon}{1 - j \tan\theta \tan\epsilon}
\end{align}
For linearly polarized light $\epsilon=0$, $\theta=\arctan\chi$ exists.


\section{Properties of attosecond light bursts in a typical interaction scenario}\label{Sec_exam}
\begin{figure}[h]\centering
\includegraphics[width=8.5cm]{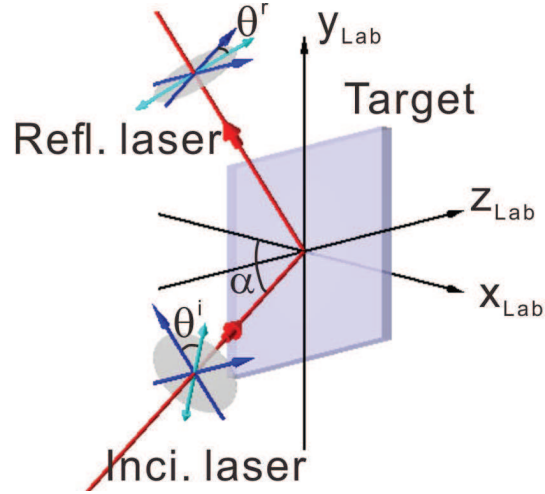}
\caption{\label{geometry} A plane wave laser pulse obliquely incident onto a plane plasma target in the lab system. Blue axes, the $p$- and the $s$- polarization axes. Cyan two-ended arrows, polarization direction or major axis of polarization ellipse. See text for symbol meanings.}
\end{figure}

As in FIG.~\ref{geometry}, we consider a plane wave laser pulse incident obliquely onto a plane plasma layer at incidence angle $\alpha$. The plasma layer has initially only one dimensional density distribution along its surface normal. The physics is Lorentz transformed into one boosted frame by using the Boudier's technique \cite{pof26.1804Bourdier83} where the problem is 1D. The boosted frame is chosen such that $x$-axis points from the laser to the plasma, perpendicular to the plasma surface and the $y$-axis lies in the plane parallel to the target surface. It has a velocity of $\mathbf{v}_0=\mathbf{\hat{y}}c\sin\alpha$ with respect to the Lab-frame, where $c$ is the vacuum light velocity. Thus, the electrons and ions initially at rest in the lab frame have the same drift velocity $-\mathbf{\hat{y}}c\sin\alpha$ in the boosted frame. $\theta$ and $\epsilon$ are invariants under the Lorentz boost. In this paper, unless otherwise stated, all the variables are from the boosted frame.

The simulation is performed using the 1D PIC code LPIC++\cite{pop3.3425Lichters96}. The incident laser pulse at $x=0$ is assumed to have a Gaussian temporal profile with the normalized electric field given by $\mathbf{E}_{\perp}^i(t)= {\rm {Re}} \{ \tilde{E}^i_{\perp}(t)\exp[j \phi^i(t)] \mathbf{\hat{e}}_{\rho}^i \}$, where $\tilde{E}^i_{\perp}(t)=E_{\rm L}\exp{[-2\ln2(t/\bar{\tau}_{\rm L})^2 ]}$ and $\phi^i(t)=2\pi t+\varphi_{\rm{CEP}}$. $E_{\rm L}$ is the peak laser field normalized by $mc\omega_{\rm L}/e$; $m$, $\omega_{\rm L}$ and $e$, the electron rest mass, laser fundamental frequency and electron charge respectively. $\bar{\tau}_{\rm L}$ is the intensity full-width-half-maximum (FWHM) laser pulse duration normalized to laser period $T_{\rm L}$. Throughout this paper we assume $E_{\rm L}\approx10$, which corresponds to a laser intensity of $2\times10^{20}~\mathrm{W/cm^{2}}$ for a laser central wavelength $\lambda_{\rm L}$ of $800~\mathrm{nm}$. The density profile of the interacting plasma has an exponential interface layer in the front with scale length of $L$. It rises from $0.2n_{c}$ up to a maximum of $90n_{c}$ and then it is followed by a $2\lambda_{\rm L}$ thick constant density distribution, where $n_e$ and $n_i$ are electron and ion fluid densities normalized by critical density $n_c$ at $\omega_{\rm L}$. A simulation box with a total length of $12\lambda_{\rm L}$
is aligned on the $x$ axis from $x=0$ to $x=12\lambda_{\rm L}$. The $2\lambda_{\rm L}$ thick plasma with flat top density is located between $x=9\lambda_{\rm L}$ and $x=11\lambda_{\rm L}$. The laser pulse is incident onto the plasma layer at  an angle of $\alpha=45^\circ$. The simulations are performed for moving ions with charge number $Z=10$ and ion to electron mass ratio $m_i/m_e=50000$. The resolution is $\Delta x=10^{-3}~\lambda_{\rm L}$ and $\Delta t=7.07\times10^{-4}~T_{\rm L}$, with 900 particles per cell for both electrons and ions at the highest density $90 n_c$.

\begin{figure}[h]\centering
\includegraphics[width=8.5cm]{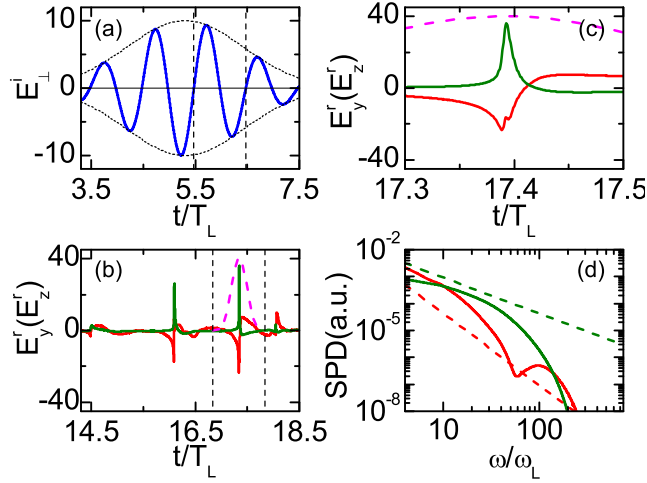}
\caption{\label{F_EyEz_SPD_t_f} (a) Incident electric field $\mathbf{E}_{\perp}^i\cdot\mathbf{\hat{e}}_{\rho}^i$ (blue solid curve). (b) Reflected electric field components. (c) A zoom in of the strongest burst shown in (b). (d) SPDs for field components of the strongest burst. Red and green solid curves are respectively for the $p$- and the $s$-polarization. The one-cycle long unit chebyshev window (magenta dashed line) used to select the strongest burst is sketched in (b) and zoomed in (c); its real amplitude is 1. Black short-dashed curves mark the incident field envelope in (a). Black dashed lines mark the cycle for the strongest attosecond burst emission in (a) and (b). Red and green dashed lines indicate the $-8/3$ and $-4/3$ roll-off scalings. Laser and plasma parameters are $E_{\rm L}=10$, $\theta^i=10^\circ$, $\varphi_{\rm{CEP}}=210^\circ$, $\tau_{\rm{FWHM}}=5~{\rm fs}$ and $L=0.43~\lambda_{\rm L}$.}
\end{figure}

The reflected laser field is written as $\mathbf{E}_{\perp}^r(t) = E_y^r(t)\mathbf{\hat{y}} + E_z^r(t)\mathbf{\hat{z}} ={\rm {Re}} \{ \tilde{E}^r_{\perp}(t) \exp[j\phi^r(t)] \mathbf{e}_{\rho}^r(t) \} $. If the incident laser pulse is $p$-polarized $\mathbf{\hat{e}}_{\rho}^i=\mathbf{\hat{y}}$ ( $\theta^i=0$, $\epsilon^i=0$, $\sigma^i=0$ and $\chi^i=0$); the reflected electric field has no $s$-polarization component. We consider here an incident pulse with $\mathbf{\hat{e}}_{\rho}^i=\cos10^\circ\mathbf{\hat{y}}+\sin10^\circ\mathbf{\hat{z}}$  ($\theta^i=10^\circ$, $\epsilon^i=0$, $\sigma^i=0$ and $\chi^i=0.18$) as in FIG.~\ref{F_EyEz_SPD_t_f}~(a), its $s$-component is $18$ percent of its $p$-component in amplitude and $3$ percent in intensity. We note that the two components of the reflected field shown in FIG.~\ref{F_EyEz_SPD_t_f}~(b) are naturally attosecond spikes even without filtering. The $s$-component spikes have larger values than their $p$-component counterparts although the $s$-portion in the incident field is small. These spikes occur at the same time in the $p$ and $s$ components, forming a train of attosecond bursts (attotrain) $\mathbf{E}^r_{\perp}(t_p)$. As an alternative, a spectral band containing the 10th to 50th laser harmonics (H10-H50) is used to synthesize a filtered field $\mathbf{E}^r_{\perp \rm fltr}(t)$. The corresponding attotrain is $\mathbf{E}^r_{\perp \rm fltr}(t_p)$.  Here $t_p$ is defined at the peaks of these bursts. Field components of the strongest burst $\mathbf{E}^r_{\perp}(t_p=t_m)$ are shown in FIG.~\ref{F_EyEz_SPD_t_f}~(c). We see $t_m\approx17.4$. Before the evaluation of the polarization state with Eq.~\ref{E_stokes} to \ref{E_polchi}, electric field of this burst is gated with a one-cycle long chebyshev window. If the gated field without filtering is evaluated, the polarization state is $\theta^r(t_p = t_m) \approx -36.4^\circ$, $\epsilon^r(t_p = t_m) \approx 17.0^\circ$, $\sigma^r(t_p = t_m) \approx 0.30$ and $\chi^r(t_p = t_m) \approx -0.64 + 0.45 j$. If the filtered field from H10-H50 is evaluated, the polarization state changes to be $\theta^r_{\rm atto}(t_p = t_m) \approx -61.0^\circ$, $\epsilon^r_{\rm atto}(t_p = t_m) \approx 15.1^\circ$, $\sigma^r_{\rm atto}(t_p = t_m) \approx 0.27$, and $\chi^r_{\rm atto}(t_p = t_m)  \approx -1.35 + 0.93j$. The corresponding polarization ellipse is approaching linear polarization and its major axis quite off the original $10^\circ$ for the incident laser pulse ($|\theta^r_{\rm atto}(t_p = t_m)| \gg |\theta^i|$). The gated field from each polarization is fourier transformed into the corresponding spectral power density (SPD) and shown in FIG.~\ref{F_EyEz_SPD_t_f}~(d). It's apparent that the spectral power law roll-off scaling of the $s$-polarized harmonics ($\sim-4/3$) is much shallower than that of the $p$-polarized harmonics ($>-8/3$) and the SPD values for both polarizations at $10th$ harmonic order are comparable. In addition, we find out that the overall $s$-polarized radiation energy is around $7$ times that from the incident pulse indicating a strong energy transfer from $p$- to $s$-polarization in the emission process.

\section{Energy coupling analysis}\label{Sec_modl}

The characteristics of the reflected field in the previous section can be well explained by the CSE model\cite{pop17.033110anderBrugge10}, in which highly dense electron nanobunches are formed and copropagate with the laser wavefront. The stored energy in these nanobunches couple efficiently to the radiative electromagnetic modes and thus generates harmonics that are phase synchronized and synthesize into single electromagnetic bursts with attosecond duration. The entire emission process happens on the sub-cycle and sub-wavelength scales of the laser.

In the CSE context, a single particle picture\cite{pop17.033110anderBrugge10,prl109.245005Mikhailova12}
is appropriate. The equations governing electrons are $\mathbf{p}_{\perp}=\mathbf{a}-\mathbf{\hat{y}}\tan\alpha$, $d_t (\mathbf{\hat{x}}p_x)=-\omega_{\rm L}E_x\mathbf{\hat{x}}-(\omega_{\rm L}/\gamma)\mathbf{p}_{\perp}\times\mathbf{B}_{\perp}$ and $d_t\gamma=-(\omega_{\rm L}/\gamma)p_xE_x-(\omega_{\rm
L}/\gamma)\mathbf{p}_{\perp}\cdot\mathbf{E}_{\perp}$ where $\mathbf{E}_{\perp}$, $\mathbf{B}_{\perp}$ and $E_x$ are electromagnetic and electrostatic fields normalized by $m c \omega_{\rm L}/e$. $\mathbf{a}$ is the vector potential normalized by $m c^2/e$ and is related to $\mathbf{E}_{\perp}$ and $\mathbf{B}_{\perp}$ through $\mathbf{E}_{\perp}=-\omega_{\rm L}^{-1}\partial_t\mathbf{a}$ and $\mathbf{B}_{\perp}=c\omega_{\rm L}^{-1}\mathbf{\hat{x}}\times\partial_x\mathbf{a}$ using Coulomb
gauge, $\mathbf{p}_{\perp}$ and $p_x$ are the transverse and $x$-component electron fluid momenta normalized by $mc$, and $\gamma$ is the relativistic factor, $\mathbf{E}_{\perp}(x,t)$ and $\mathbf{B}_{\perp}(x,t)$ can be obtained from the simulation. They are separable into the forward (positive $x$) and backward (negative $x$) propagating fields using the spectral method, $\mathbf{E}_{\perp}=\mathbf{E}^f_{\perp}+\mathbf{E}^b_{\perp}$ and $\mathbf{B}_{\perp}=\mathbf{B}^f_{\perp}+\mathbf{B}^b_{\perp}$. Correspondingly, $\mathbf{a}=\mathbf{a}^f+\mathbf{a}^b$ and $\mathbf{p}_{\perp}=\mathbf{p}^f_{\perp}+\mathbf{p}^b_{\perp}-\mathbf{\hat{y}}\tan\alpha$, where $\mathbf{p}^f_{\perp}$ and $\mathbf{p}^b_{\perp}$ are the quiver momenta due to the forward and backward propagating fields. Moreover, $\mathbf{E}^f_{\perp}(x,t)$ and $\mathbf{E}^b_{\perp}(x,t)$ evolve inside plasma during the emission and are thus different from $\mathbf{E}^i_{\perp}(x,t)$ and $\mathbf{E}^r_{\perp}(x,t)$ which are assumed to freely propagate in infinite space without damping.

\begin{figure}[h]\centering
\includegraphics[width=8.5cm]{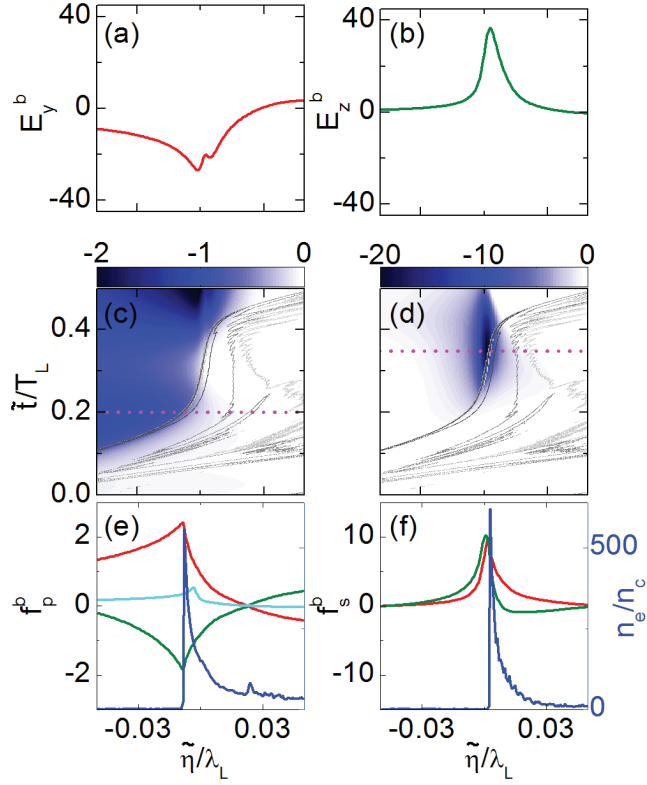}
\caption{\label{F_coupling} Emission of the strongest attosecond burst. (a)~$E^b_y(\tilde{\eta},\tilde{t}=0.5T_{\rm L})$. (b)~$E^b_z(\tilde{\eta},\tilde{t}=0.5T_{\rm L})$. (c)~$W^b_p(\tilde{\eta},\tilde{t})$. (d)~$W^b_s(\tilde{\eta},\tilde{t})$. (e)~$f^b_p(\tilde{\eta},\tilde{t}=\tilde{t}_1)$ and $n_e(\tilde{\eta},\tilde{t}=\tilde{t}_1)$ (blue solid curve). (f)~$f^b_p(\tilde{\eta},\tilde{t}=\tilde{t}_2)$ and $n_e(\tilde{\eta},\tilde{t}=\tilde{t}_2)$ (blue solid curve). Red,
green and cyan lines in (e) are the forces $f^{\rm cro}_p$, $f^{\rm sqr}_p$ and $f^{\rm lin}_p$. Red and green lines in (f) are the forces $f^{\rm cro}_s$ and $f^{\rm sqr}_s$. Magenta short dashed-lines in (c) and (d) are $\tilde{t}=\tilde{t}_1$ and $\tilde{t}=\tilde{t}_2$, when the H10-H50 harmonic energy growth rate is maximum for each polarization. $n_e(\tilde{\eta},\tilde{t})$ is shown in (c) and (d) as contours. The dark gray, gray and light gray level curves correpond to the density of $250n_c$, $60n_c$ and $30n_c$ respectively. The origins of $\tilde{\eta}$ and $\tilde{t}$ axes have been changed for appropriate presentation. Laser and plasma parameters are the same as in FIG.~\ref{F_EyEz_SPD_t_f}.}
\end{figure}

The force responsible for the emission of the reflected field in the equation for $d_t(\mathbf{\hat{x}} p_x)$ can be written as $\mathbf{f}^b=\mathbf{f}^b_{\rm elec}+\mathbf{f}^b_{\rm pond}$ with $\mathbf{f}^b_{\rm elec}=(\omega_{\rm L}/\gamma)\tan\alpha\mathbf{\hat{y}}\times\mathbf{B}^b_{\perp}$ and $\mathbf{f}^b_{\rm pond}=-(\omega_{\rm L}/\gamma)(\mathbf{p}^f_{\perp}+\mathbf{p}^b_{\perp})\times\mathbf{B}^b_{\perp}$. Using the approximation $\mathbf{E}^b_{\perp}\approx\mathbf{\hat{x}}\times\mathbf{B}^b_{\perp}$ and $p_x/\gamma\approx-1$, we have $d_t\gamma\approx-(\omega_{\rm
L}/\gamma)p_x E_x-(\omega_{\rm L}/\gamma)\mathbf{p}_{\perp}\cdot\mathbf{E}^f_{\perp}+\mathbf{f}^b_{\rm elec}\cdot(p_x/\gamma)\mathbf{\hat{x}}+\mathbf{f}^b_{\rm pond}\cdot
(p_x/\gamma)\mathbf{\hat{x}}$ with the third and fourth terms the rate of energy exchange from backward radiation to electrons by the direct electric force and the ponderomotive force respectively. The latter is well-known for its role in secular energy exchange between the electrons and the laser light. When the ponderomotive force is in the direction of electron motion, the electrons get energy from the laser field and light is absorbed and when it is in the opposite direction, the electrons lose energy to the laser field and the light is amplified.

For the emission of the strongest attosecond burst, the interaction process is best presented in a coordinate system moving with the wave front of the reflected field $(\tilde{\eta}=x+ct,\tilde{t}=t)$. We define the normalized force for $p$-polarization: $f^b_p=f^{\rm lin}_p+f^{\rm cro}_p + f^{\rm sqr}_p$ with $f^{\rm lin}_p=\tan\alpha\times B^b_z/E^2_{y\rm L}$, $f^{\rm cro}_p=- p^f_y\times B^b_z/E^2_{y\rm L}$ and $f^{\rm sqr}_p=- p^b_y\times B^b_z/E^2_{y\rm L}$; for $s$-polarization $f^b_s=f^{\rm cro}_s + f^{\rm sqr}_s$ with $f^{\rm cro}_s=p^f_z \times B^b_y/E^2_{z\rm L}$ and $f^{\rm sqr}_s=p^b_z\times B^b_y/E^2_{z\rm L}$. At a time moment when the growth rate of the harmonic energy for each polarization component within H10-H50 is around maximum, the electron nanobunch is moving at close to the light velocity $c p_x/\gamma\sim -c$. In FIG.~\ref{F_coupling}~(e) and (f), we have compared these forces with the instantaneous electron density distribution. $f^{\rm lin}_p$ has negligible magnitude compared to all other terms confirming that the pure density coupling\cite{pre82.056410Tarasevitch10} is negligible. $f^{\rm cro}_p$ is pointing in positive-$x$ direction, while $f^{\rm sqr}_p$ is pointing in negative-$x$ direction. Their effects cancel each other resulting in a low gain. However, both $f^{\rm cro}_s$ and $f^{\rm sqr}_s$ are pointing in positive $x$-direction. Normalized terms of $f^b_p$ are all weaker than those of $f^b_s$. This point favors the rotation of the major axis of pollarization ellipse towards the $s$-polarization direction. The direction of $f^{\rm sqr}_p$ or $f^{\rm sqr}_s$ determines the nature of the feedback to the emission process. The fact that $f^{\rm sqr}_s$ pointing in the opposite direction of electron motion suggests a positive feedback to the emission process, which also happens in high gain regime of FELs. This also indicates a high coherence factor\cite{springer9783540306900Buts06} for the electron bunch in emitting the $s$-polarized light.

Corresponding to $f^b_p$ and $f^b_s$, we define the normalized energy exchange rate for $p$- and $s$-polarized backward radiation as $W^b_p= - {p_y E^b_y}/{E^2_{y\rm L}}$ and $W^b_s=- {p_z E^b_z}/{E^2_{z\rm L}}$. The evolution of $W^b_p$ and $W^b_s$ for the emission of the strongest attosecond burst are compared in FIG.~\ref{F_coupling}~(c) and (d). Their negative values mean that the energy exchanges from electrons to radiation field. At a fixed $\tilde{\eta}$, $W^b_p(\tilde{\eta},\tilde{t})$ and $W^b_s(\tilde{\eta},\tilde{t})$ are integrable along $\tilde{t}$ to give the total energy accumulation at that $\tilde{\eta}$. Due
to energy exchanges, the magnitudes of backward radiation fields $E^b_y(\tilde{\eta},\tilde{t}=0)$ and $E^b_z(\tilde{\eta},\tilde{t}=0)$ grow from trivial values to a state at the end of radiation process as shown in FIG.~\ref{F_coupling}~(a) and (b). After $\tilde{t}=0.5T_{\rm L}$, the interaction is over and the shapes of $E^b_y$ and $E^b_z$ don't change anymore. They are observed at simulation box end as $E_y^r(t)$ and $E_z^r(t)$. We see that $W^b_p$ is dispersed and
weak while $W^b_s$ is localized and strong. Besides, energetic high density electrons overlapped with $W^b_s$ much better than with $W^b_p$. This signifies that energy extraction from electrons to $s$-polarization field favors shorter spike and higher conversion efficiency. We also see that the width of $E^b_z(\tilde{\eta},\tilde{t}=0.5T_{\rm L})$ is narrower than $E^b_y(\tilde{\eta},\tilde{t}=0.5T_{\rm L})$, in agreement with the strong and shallow spectrum for $E^r_z$ shown in
FIG.~\ref{F_EyEz_SPD_t_f}~(d). Detailed investigation shows that it is the strong nonsymmetric depletion of $\mathbf{E}^i_{\perp}$ to $\mathbf{E}^f_{\perp}$ by the cloud electrons from the previous cycle before $\mathbf{E}^f_{\perp}$ sees the bunch and by the bunch during the emission process which makes $\mathbf{\hat{e}}^f_\rho(\tilde{\eta},\tilde{t})\neq \mathbf{\hat{e}}^i_\rho$ and $E^f_y$ very different from $E^f_z$ in shape. Here $\mathbf{\hat{e}}^f_\rho$ is the polarization vector of $\mathbf{E}^f_\perp$. This resulted in a higher gain for the $s$-polarized light which makes $|\chi^r| \gg |\chi^i|$, and the rotation of the major axis of the polarization ellipse of an attosecond burst towards the $s$-polarization direction.

 Our analysis here can be compared with that of Tarasevitch {\it et al.}\cite{jpbamop42.134006Tarasevitch09}. They proposed a two-beam HHG concept and considered the energy exchange between the two superposed beams at different laser frequencies. And they used an externally driven oscillating mirror model\cite{jpbamop42.134006Tarasevitch09} to explain the much shallower spectral shape for the reflected weak probe beam and later they identified two different energy coupling mechanisms for the two beam HHG:\cite{pre82.056410Tarasevitch10} ``pure density'' and ``density-velocity'' coupling, with the pure density coupling prevalent in the CWE dominated regime. The existent models discuss harmonic emission due to the ROM or the CWE mechanism. Here we consider CSE harmonic emission; the role of ponderomotive force in the energy coupling process is emphasized.

\section{Parametric dependence}\label{Sec_para}
\begin{figure}[t]\centering
\includegraphics[width=8.5cm]{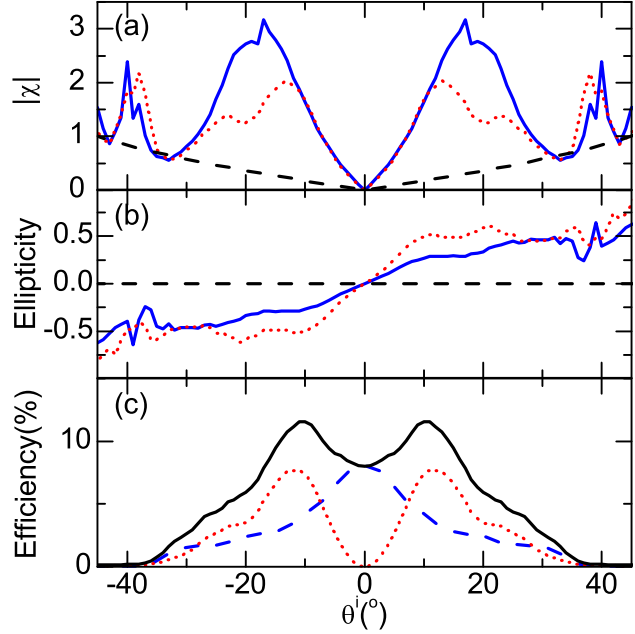}
\caption{\label{F_harmpol_azimuth}(a) $\theta^i$ dependence of (a) $|\chi|$ and (b) $\sigma$ for the strongest attosecond burst (blue solid curve) and for the attotrain (red dotted line) synthesized from H10-H50. Black dashed curve, $\chi^i$ in (a) and $\sigma^i$ in (b) of the incident laser pulse as a reference. (c) $\theta^i$ dependence of XUV conversion efficiency for H10-H50. Blue dashed, red dotted and black solid curves in (c) correspond to the $p$-polarized $\eta_p$, the $s$-polarized $\eta_s$ and the overall $\eta_t$ harmonic conversion efficiencies. The scan resolution for $\theta^i$ is $1^\circ$. Other laser and plasma parameters are $E_{\rm L}=10$, $\varphi_{\rm{CEP}}=210^\circ$, $\tau_{\rm{FWHM}} =5~{\rm fs}$ and $L=0.43\lambda_{\rm L}$. }
\end{figure}

 Among the many possible combinations of $\theta^i$ and $\epsilon^i$ for the incident laser pulse, we concentrate on two special cases: 1) when $\epsilon^i=0$, $\mathbf{\hat{e}}_{\rho}^i=\cos\theta^i \mathbf{\hat{y}} + \sin\theta^i \mathbf{\hat{z}}$, the pulse is linearly polarized with azimuth $\theta^i$. 2) when $\theta^i=0$, $\mathbf{\hat{e}}_{\rho}^i = \cos\epsilon^i \mathbf{\hat{y}} + j \sin\epsilon^i \mathbf{\hat{z}}$, the pulse is elliptically polarized with major axis along $y$-axis and ellipticity $\sigma^i = \tan\epsilon^i$. The filtered field $\mathbf{E}^r_{\perp \rm fltr}$ from spectral band H10-H50 is used to evaluate the polarization state of a single attosecond burst or an attotrain. The polarization state for the strongest attosecond burst is described by $\theta^r_{\rm atto}$, $\epsilon^r_{\rm atto}$, $\sigma^r_{\rm atto}$ and $\chi^r_{\rm atto}$. The polarization state for the whole attotrain is described by $\theta^r_{\rm train}$, $\epsilon^r_{\rm train}$, $\sigma^r_{\rm train}$ and $\chi^r_{\rm train}$. And we define the XUV conversion efficiencies as $\eta_p = \int ( \mathbf{\hat{y}} \cdot \mathbf{E}^r_{\perp \rm fltr} )^2 dt / \int \vert \mathbf{E}_{\perp}^i \vert^2 dt$ for the $p$-polarized, $\eta_s = \int ( \mathbf{\hat{z}} \cdot \mathbf{E}^r_{\perp \rm fltr} )^2 dt / \int \vert \mathbf{E}_{\perp}^i \vert^2 dt$ for the $s$-polarized, and $\eta_t = \eta_p + \eta_s$ for the total.

When the azimuth $\theta^i$ of a linearly polarized laser pulse is varied, we have made a series of PIC simulations and have got $|\chi^r_{\rm atto}|$ and $|\chi^r_{\rm train}|$, as shown in FIG.~\ref{F_harmpol_azimuth}~(a); as well as  $\sigma^r_{\rm atto}$ and $\sigma^r_{\rm train}$, as shown in FIG.~\ref{F_harmpol_azimuth}~(b). The variation of $|\chi^r_{\rm atto}|$ or $|\chi^r_{\rm train}|$ is much faster than the variation of $|\chi^i|$. (When $\sigma\sim0$, the fast change of $|\chi|$ indicates a fast rotation of $\theta$.) When $|\chi^i|$ linearly increases up to
$|\chi^i|\approx0.18$ ($|\theta^i|=10^\circ$), $|\chi^r_{\rm atto}|$ and $|\chi^r_{\rm train}|$ almost linearly increase up to $|\chi^r_{\rm atto}|\approx1.64$ and $|\chi^r_{\rm train}|\approx1.68$ accordingly. At $\theta^i>10^\circ$, there exists a large $\theta^i$ range where $|\chi^r_{\rm atto}|$ is much higher than $|\chi^r_{\rm train}|$. Although the incident laser pulse is linearly polarized, the attosecond bursts generally show certain degree of ellipticity, however they still approach linear polarization for small $\theta^i$s. The XUV conversion efficiencies are shown in FIG.~\ref{F_harmpol_azimuth}~(c). We see that all the curves are symmetric about $\theta^i=0$. When the incident pulse is purely $p$-polarized $\theta^i=0$, $|\chi^r_{\rm atto}|=|\chi^r_{\rm train}|\equiv0$. When it is rotated off $p$-polarization, the $p$-polarized harmonic energy decrease monotonically while the $s$-polarized one increases first and then decreases again. For $\vert\theta^i\vert>45^\circ$, $\eta_p$, $\eta_s$ and $\eta_t$ are orders of magnitude lower than $\eta_p(\theta^i=0)$ and thus not of interest.  There exists an optimum $\theta^i$ for the incident laser pulse $\vert\theta^i_{\rm opt}\vert\approx10^\circ$ where both $\eta_s$ and $\eta_t$ are maximized. Although $\theta^i_{\rm opt}$ is not large, $\eta_s$ around $\theta^i_{\rm opt}$ holds a much larger portion than $\eta_p$ in the overall efficiency $\eta_t$.

\begin{figure}[h]\centering
\includegraphics[width=8.5cm]{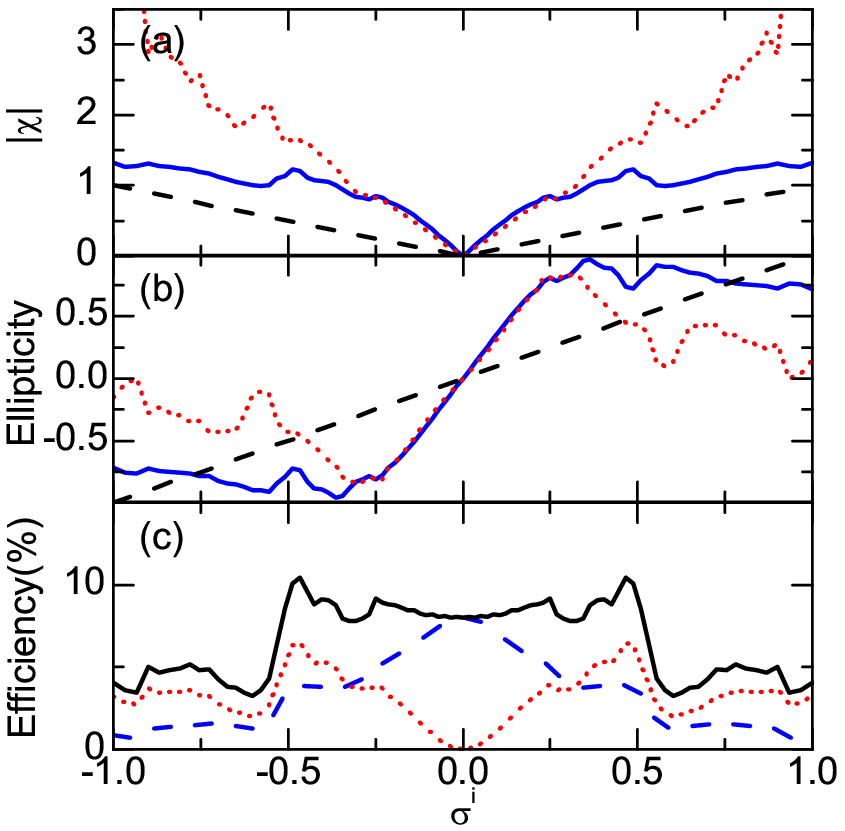}
\caption{\label{F_harmpol_sigma}(a) $\sigma^i$ dependence of (a) $|\chi|$ and (b) $\sigma$ for the strongest attosecond burst (blue solid curve) and for the attotrain (red dotted line) synthesized from H10-H50. Black dashed curve, $\chi^i$ in (a) and $\sigma^i$ in (b) of the incident laser pulse as a reference. (c) $\sigma^i$ dependence of XUV conversion efficiency for H10-H50. Blue dashed, red dotted and black solid curves in (c) correspond to the $p$-polarized $\eta_p$, the $s$-polarized $\eta_s$ and the overall $\eta_t$ harmonic conversion efficiencies. $\epsilon^i$ is scanned at a resolution of $1^\circ$ in the range $-45^\circ\leq\epsilon^i\leq45^\circ$ to give $\sigma^i=\tan\epsilon^i$. Other laser and plasma parameters are $E_{\rm L}=10$, $\varphi_{\rm{CEP}}=210^\circ$, $\tau_{\rm{FWHM}} =5~{\rm fs}$ and $L=0.43\lambda_{\rm L}$.}
\end{figure}

\begin{figure}[h]\centering
\includegraphics[width=8.5cm]{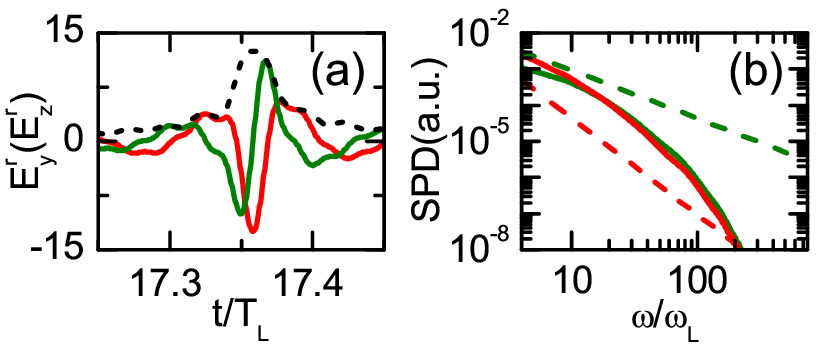}
\caption{\label{F_circatto_tf} Time and spectrum structures of the strongest attosecond burst. (a) red solid, green solid and black short dashed lines are for $E^r_{y,\rm atto}$, $E^r_{z,\rm atto}$ and $[(E^r_{y,\rm atto})^2 + (E^r_{z,\rm atto})^2]^{1/2}$ respectively. Spectral range H10-H50 is used to synthesize the burst. (b) red and green solid lines are SPDs for $p$- and $s$-polarized light correspondingly. Red and green dashed lines are the $-8/3$ and $-4/3$ power law roll-off scalings. Laser and plasma parameters are $E_{\rm L}=10$, $\varphi_{\rm CEP}=210^\circ$, $\tau_{\rm FWHM}=5~{\rm fs}$, $\rm L = 0.43 \lambda_{\rm L}$ and $\epsilon^i=20^\circ$.}
\end{figure}

When $\theta^i=0$ and the ellipticity of the incident laser pulse $\sigma^i=\tan\epsilon^i$ is varied, the PIC simulation results for $|\chi^r_{\rm atto}|$ and $|\chi^r_{\rm train}|$ are shown in FIG.~\ref{F_harmpol_sigma}~(a); while $\sigma^r_{\rm atto}$ and $\sigma^r_{\rm train}$ are shown in FIG.~\ref{F_harmpol_sigma}~(b). The variation of $|\chi^r_{\rm atto}|$ and $|\chi^r_{\rm train}|$ is still faster than the variation of $|\chi^i|$. The variation of $|\sigma^r_{\rm atto}|$ or $|\sigma^r_{\rm train}|$ is much faster than the variation of $|\sigma^i|$. When $|\sigma^i|$ linearly increases up to $|\sigma^i|\approx0.25$ ($|\epsilon^i|=14^\circ$), $|\sigma^r_{\rm atto}|$ and $|\sigma^r_{\rm train}|$ almost linearly increase up to $|\sigma^r_{\rm atto}=|\sigma^r_{\rm train}|\approx0.81$ ($|\epsilon^r_{\rm atto}=|\epsilon^r_{\rm train}|\approx39^\circ$) accordingly. When the incident pulse has an ellipticity of $|\sigma^i|\approx0.36$ ($|\epsilon^i|\approx20^\circ$), the ellipticity of the strongest attosecond burst is even as high as $\sigma^r_{\rm atto}\approx0.95$ ($|\epsilon^r_{\rm atto}\approx43.6^\circ|$). The temporal structures as well as the SPDs for this burst is shown in FIG.~\ref{F_circatto_tf}~(a) and (b). For elliptically polarized light with small ellipticity, the generated attosecond burst is almost circularly polarized. The XUV conversion efficiencies are shown in FIG.~\ref{F_harmpol_sigma}~(c). We see that all the curves are symmetric about $\sigma^i=0$. When the incident pulse is purely $p$-polarized $\sigma^i=0$, $|\sigma^r_{\rm atto}|=|\sigma^r_{\rm train}|\equiv0$. When it is rotated off $p$-polarization, the $p$-polarized harmonic energy decrease monotonically while the $s$-polarized one increases first and then decreases again. There exists a $\sigma^i$ range where the overall efficiency $\eta_t$ almost does not change. We note that, for the case where the generated attosecond burst is circularly polarized, the XUV conversion efficiency is comparable to that of a $p$-polarized driver ($\theta^i=\epsilon^i=0$). Although in principle, a normally incident circularly polarized laser can also generate a circularly polarized attosecond burst\cite{pop18.083104Ji11}. However, the requirements on the driving laser pulse is  demanding and the XUV conversion efficiency is much lower than the corresponding oblique incidence case.

\begin{figure}[h]\centering
\includegraphics[width=8.5cm]{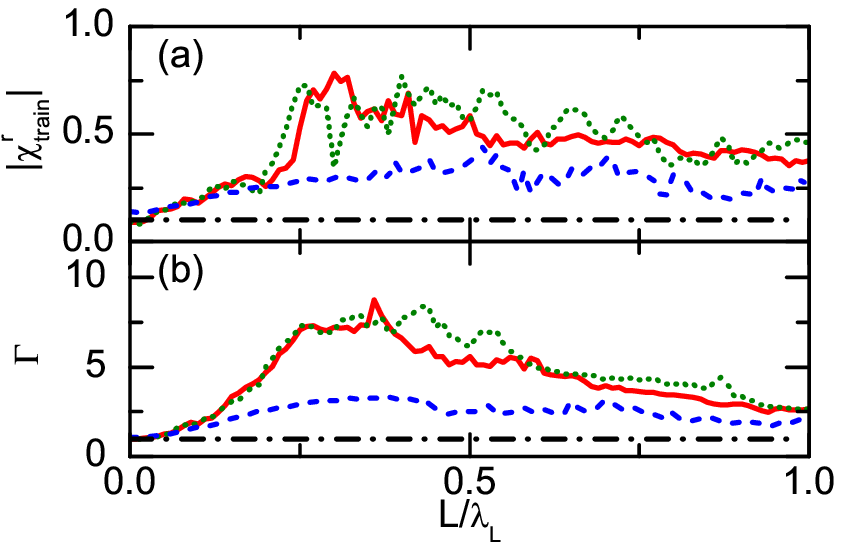}
\caption{\label{F_harmengamp_L} $L$ dependence of (a) $|\chi|$ for the attotrain synthesized from H10-H50, and of (b) the ratio $\Gamma$ for $s$-polarized light for the whole spectral range. Red solid curve, $\varphi_{\rm{CEP}}=135^\circ$ and $\tau_{\rm{FWHM}}=5~{\rm fs}$; green short-dotted curve, $\varphi_{\rm{CEP}}=210^\circ$ and $\tau_{\rm{FWHM}}=5~{\rm fs}$; blue short-dashed curve, $\varphi_{\rm{CEP}}=0^\circ$ and $\tau_{\rm FWHM}=25~{\rm fs}$.
The scan resolution for $L$ is $0.01\lambda_{\rm L}$. Laser amplitudes are fixed at $E_{y\rm L}=10$ and $E_{z\rm L}=1$ with polarization state $\theta^i=5.7^\circ$ and $\epsilon^i=0$. Black dash-dotted line in (a) $|\chi_{\rm train}^r| = 0.1$, in (b) $\Gamma=1$.}
\end{figure}

It is known that the carrier envelope phase (CEP), the driver pulse duration, and the plasma scale length can all affect the efficiency of harmonic generation\cite{pop22.033105Ma15} In FIG.~\ref{F_harmengamp_L}, we compare the effects of these parameters for the laser field amplitudes $E_{y\rm L}=10$ and $E_{z\rm L}=1$. As in (a), for $\tau_{\rm FWHM}=5~{\rm fs}$, at the scale length $L_{\rm opt}\approx0.36\lambda_{\rm L}$, $|\chi^r_{\rm train}|$ is as large as $|\chi^r_{\rm train}|\approx 0.62$, compared with the small $\chi^i$ of the incident pulse $\chi^i=0.1$. The $L$ dependence of $|\chi^r_{\rm train}|$ is almost unaffected by $\varphi_{\rm CEP}$ but is affected by $\tau_{\rm FWHM}$. For $\tau_{\rm FWHM}=25~{\rm fs}$, $|\chi^r_{\rm train}|$ decreases to $|\chi^r_{\rm train}|\approx0.32$ at optimum scale length. FIG.~\ref{F_harmengamp_L}~(b) shows the energy amplification ratio $\Gamma$ for the $s$-polarized electromagnetic field. Here $\Gamma$ is defined as $\Gamma=\int (E^r_z)^2 dt/\int (E^i_z)^2 dt$. An energy amplification as high as $\Gamma\approx8.7$ has been found for $5~{\rm fs}$ driver pulse at optimum scale length for both CEPs. For a longer driver pulse where $\tau_{\rm FWHM}=25~{\rm fs}$, the CSE is not the dominated HHG process\cite{pop22.033105Ma15}; it experiences a weaker amplification $\Gamma\approx3.3$ at optimum. This energy amplification is a clear evidence of energy exchange from the streaming electrons to the $s$-component of the reflected laser field. The calculation is repeated for $E_{y{\rm L}}=10$ and $E_{z{\rm L}}=0.1$. In this case all the curves shown almost retain their shapes, although the $|\chi^r_{\rm train}|$ values in (a) decrease.

\section{discussions and conclusions}\label{Sec_conc}
To conclude, we have investigated the polarization properties of attosecond light bursts as well as its parametric dependence, for the CSE harmonic generation in the ``oblique incidence in reflection'' geometry. A detailed energy coupling analysis is done to reveal the underlying physics: the effect of ponderomotive force on electron nanobunches determine the polarization state of the generated attosecond light bursts.

Our results help understanding the characteristics of laser-plasma based attosecond light sources, especially for diagnosing and optimizing the interaction parameters for polarization controlled isolated ultrabright attosecond light bursts.

Based on the results, we propose that, if the laser and plasma parameters (pulse duration, carrier envelope phase, plasma scale length) are controlled, through the manipulation of the polarization state of the driver laser pulse, we can generate an intense circularly polarized attosecond light burst from plasma surfaces which would benefit to many advanced applications.
 
Due to the computation complexity, the analysis done in this work is limited to situation convertible into 1D problems. However, the underlying physics is so enlightening that we believe, although detailed dependencies are changable due to multi-dimensional effects in real experiments, we can still expect the same conclusion.


\section*{Acknowledgements}

The work was supported by the National Natural Science Foundation of China (11304331, 11174303, 11374262), the National Basic Research Program of China (2013CBA01504), the Fundamental Research Funds for Central Universities, the Munich Center for Advanced Photonics(MAP), DFG Project Transregio TR18, and the Association EURATOM Max-Planck-Institut f\"ur Plasmaphysik.

\end{document}